\documentclass[12pt]{article}
\usepackage[dvips]{color}
\usepackage{epsfig}
\usepackage{amsmath}
\usepackage{graphicx}

\begin{document}
\title{\bf Phenomenologically varying $\Lambda$ and a toy model for the Universe}
\author{ {M. Khurshudyan$^{a}$ \thanks{Email: khurshudyan@yandex.ru},
\hspace{1mm} J. Sadeghi$^{b}$ \thanks{Email: pouriya@ipm.ir},\hspace{1mm}\hspace{1mm} E. Chubaryan$^{a}$ \thanks{Email: echub@ysu.am}
and H. Farahani$^{b}$ \thanks{Email: h.farahani@umz.ac.ir}}\\
$^{a}${\small {\em Department of Theoretical Physics, Yerevan State
University,}}\\
{\small {\em 1 Alex Manookian, 0025, Yerevan, Armenia}}\\
$^{b}${\small {\em Department of Physics, Ayatollah Amoli Branch, Islamic Azad University, Amol, Iran}}\\
{\small {\em P.O.Box 678, Amol, Iran}}}  \maketitle
\begin{abstract}
We consider a model of the Universe with variable $G$ and $\Lambda$. Subject of our interest is a phenomenological model for $\Lambda$ proposed and considered in this article first time (up to our knowledge). Modification based on an assumption that ghost dark energy exists and Universe will feel it through $\Lambda$. In that case we would like to consider possibility that there exist some unusual connections between different components of the fluids existing in Universe. We would like to stress, that this is just an assumption and could be very far from the reality. We are interested by this model as a phenomenological and mathematical and unfortunately, we will not discuss about physical conditions and possibilities of having such modifications. To test our assumption and to observe behavior of the Universe, we will consider toy models filled by a barotropic fluid and modified Chaplyagin gas. To complete the logic of the research we will consider interaction between barotropic fluid or Chaplygin gas with ghost dark energy as a separate scenario. Statefinder diagnostic also provided with stability analysis of the models. All free parameters of the model fixed to satisfy generalized second law of thermodynamics.
{\bf Pacs Number(s):} 95.35.+d, 95.85.-e, 98.80.-k, 95.30.Cq, 97.20.Vs, 98.80.Cq
\end{abstract}
\section*{\large{Introduction}}
Experimental data and observations claim that our Universe undergoes accelerated expansion \cite{Riess}-\cite{Amanullah}. But we should remember that this conclusion depends on theoretical model under consideration. In general relativity to explain accelerated expansion a dark energy concept was introduced by hand. Dark energy is a fluid with negative pressure and positive energy density giving negative EoS parameter. Different ways are used in order to understand the "origin" of dark energy. It is argued that the dark energy is about $73\%$ of the energy of our Universe. Other component, dark matter occupies about $23\%$, and usual baryonic matter occupies about $4\%$. Some authors thought that it could be a scalar field, which is accepted and used approach in cosmology. A quintessence field is a possible model based on scalar field with standard kinetic term, which minimally coupled to gravity. In that case the action has a wrong sign kinetic term and the scalar field is called phantom. Combination of the quintessence and the phantom is known as the quintom model. Extension of kinetic term in Lagrangian yields to a more general frame work on field theoretic
dark energy, which is called k-essense. A singular limit of k-essense is called Cuscuton model.
This model has an infinite propagating speed for linear perturbations, however causality is still valid. The
most general form for a scalar field with second order equation of motion is the Galileon field which also
could behaves as dark energy \cite{Wetterich}-\cite{Murli}. One of the ways is to take geometrical part of the action and modify, which results different modified versions of gravity like to $F(R)$, $F(G)$, $F(R,T)$, $F(T)$ e.t.c \cite{Odsintsov}-\cite{Saridakis}. Modifications of this type provide an origin of a fluid (energy density and pressure) identified with dark energy. This fluids obviously have geometrical origin. A possibility to have accelerated expansion contributed from geometry were considered even before proposed modifications and this approach could be accepted as true-like feeling. But such theories with different forms of modifications (phenomenological) still should pass experimental tests, because they contain ghosts, finite-time future singularities e.t.c, which are very troubled and we should be sure that they are just a result of not "correct" form of modification. Actually the number of good articles and educational materials are so much that it is very hard to mention one or several of them to readers. We make citations to some of several researchers only and references therein could be very useful. Other approach also exists and is considered intensively in literature related to the energy source existing in field equations. The form of the $\rho$ on right hand side of field equations is an average quantity which depend on scales and not always it could be used as we use. This is one of the reasons to consider inhomogeneous fluids. For more critics and reasoning concerning to the subject can be found, for instance in \cite{Chia} and references therein. We can discuss a lot about possible holes and counterintuitive results and scenarios in theories for Universe, we can develop different philosophy e.t.c, but the goal of this article is different. We would like to consider a phenomenological model for $\Lambda$, which will contain information about one of the models developed for dark energy called ghost dark energy (GDE). As a toy model of The Universe we will consider two different cases.\\
We are interested by the Universe with varying $G$ and $\Lambda$ which contains,
\begin{enumerate}
\item Barotropic fluid with $P_{b}=\omega \rho_{b}^{n}$ ($n > 0$) [19],
\item  modified Chaplygin gas with $P_{MCG}=A\rho_{MCG}-\frac{B}{\rho_{MCG}^{n}}$ ($ 0 \le n <1 $) [20-24],
\end{enumerate}
as a model. Second model of our interest is to consider two-fluid component Universe, where Barotropic fluid and Chaplygin gas interacts with GDE. Concluding this section we would like to summarize the goal of the article. The goal is to consider a possibility to have unusual connections between fluid components i.e. non apparent and usual contribution of fluids into the dynamics of Universe as was considered before in literature. Our assumption is that according to some physics some fluids will appear into field equations and dynamics with, for instance in our case with varying $\Lambda$. Other possibility is to have non usual interaction between fluid components containing some information about other fluids.\\
The paper organized as follow: In the next section we discuss field equations in case of variable $G(t)$ and $\Lambda(t)$, metric of our Universe as well as origin of GDE and a form for $\Lambda(t)$. Then, in other sections we will investigate behavior of cosmological parameters like deceleration parameter $q$, EoS parameter $\omega_{tot}$ etc for both models. Stability of the models is also considered. Finally, some conclusion and discussions are provided for the models.

\section*{\large{FRW Universe with variable $G$ and $\Lambda$}}
In this article we consider a possibility, that our Universe has GDE \cite{Ghost1}-\cite{Chao-Jun}, which is a model of dark energy supposed to exist to solve the $U(1)_{A}$ problem in low-energy effective theory of QCD. It has attracted a lot of interests in
recent years. Indeed, the contribution of the ghosts field to the vacuum energy in curved space or time-dependent background can be regarded
as a possible candidate for the dark energy. It is completely decoupled from the physics sector. Veneziano ghost is unphysical in the QFT formulation in Minkowski space-time, but exhibits important non trivial physical effects in the expanding Universe and these effects give rise to a vacuum energy density $\rho_{D}\sim \Lambda^{3}_{QCD}H\sim (10^{-3}eV)^{4}$. With $H\sim 10^{-33}eV$ and $\Lambda_{QCD}\sim 217 MeV$ we have the right value for the force accelerating the Universe today. It is hard to accept such linear behavior and it is thought that there should be some exponentially small corrections. However, it can be argued that the form of this behavior can be result of the fact of the very complicated topological structure of strongly coupled CQD. This model has advantage compared to other models of dark energy, that it can be explained by standard model and general relativity. Comparison with experimental data reveal that the current data does not favorite compared to the $\Lambda$CDM model, which is not conclusive and future study of the problem is needed. Energy density of GDE reads as,
\begin{equation}\label{eq:GDE}
\rho_{\small{GDE}}=\theta H,
\end{equation}
where $H$ is Hubble parameter $H=\dot{a}/a$ and $\theta$ is constant parameter of the model, which should be determined. A generalization of the model \cite{Cai} also was proposed for which energy density and reads as,
\begin{equation}\label{eq:GDEgen}
\rho_{\small{GDE}}=\theta H+\beta H^{2},
\end{equation}
with $\theta$ and $\beta$ constant parameters of the model. Recently, idea of varying GDE \cite{Khurshudyans} were proposed, which from our opinion still should pass long way with different modifications.
At the same time we propose to consider a phenomenological model for a varying $\Lambda$ of the form,
\begin{equation}\label{eq:lambda}
\Lambda(t)=\frac{\rho^{2}}{\rho+\alpha\rho_{\small{GDE}}}+\rho_{GDE}\exp(-tH),
\end{equation}
which is a key assumption of recent investigation.
Field equations that govern our model with variable $G(t)$ and
$\Lambda(t)$ (see for instance \cite{Abdussattar}, \cite{Ujjal}) are,
\begin{equation}\label{s1}
R^{ij}-\frac{1}{2}Rg^{ij}=-8 \pi G(t) \left[ T^{ij} -
\frac{\Lambda(t)}{8 \pi G(t)}g^{ij} \right],
\end{equation}
where $G(t)$ and $\Lambda(t)$ are function of time. By using the
following FRW metric for a flat Universe,
\begin{equation}\label{s2}
ds^2=-dt^2+a(t)^2\left(dr^{2}+r^{2}d\Omega^{2}\right),
\end{equation}
field equations can be reduced to the following Friedmann equations,
\begin{equation}\label{eq: Fridmman vlambda}
H^{2}=\frac{\dot{a}^{2}}{a^{2}}=\frac{8\pi G(t)\rho}{3}+\frac{\Lambda(t)}{3},
\end{equation}
and,
\begin{equation}\label{eq:fridman2}
\frac{\ddot{a}}{a}=-\frac{4\pi
G(t)}{3}(\rho+3P)+\frac{\Lambda(t)}{3},
\end{equation}
where $d\Omega^{2}=d\theta^{2}+\sin^{2}\theta d\phi^{2}$, and $a(t)$
represents the scale factor. The $\theta$ and $\phi$ parameters are
the usual azimuthal and polar angles of spherical coordinates, with
$0\leq\theta\leq\pi$ and $0\leq\phi<2\pi$. The coordinates ($t, r,
\theta, \phi$) are called co-moving coordinates.\\
Energy conservation $T^{;j}_{ij}=0$ reads as,
\begin{equation}\label{eq:conservation}
\dot{\rho}+3H(\rho+P)=0.
\end{equation}
Combination of (\ref{eq: Fridmman vlambda}), (\ref{eq:fridman2}) and (\ref{eq:conservation}) gives the relationship between $\dot{G}(t)$ and $\dot{\Lambda}(t)$ as the following,
\begin{equation}\label{eq:glambda}
\dot{G}=-\frac{\dot{\Lambda}}{8\pi\rho}.
\end{equation}
Ever since Dirac's proposition of a possible time variation of $G$,
a volume of works has been centered around the act of calculating
the amount of variation of the gravitational constant. For instance,
observation of spinning-down rate of pulsar PSR J2019+2425
provides the result,
\begin{equation}\label{eq:gvar1}
\left|\frac{\dot{G}}{G} \right|\leq (1.4-3.2) \times 10^{-11} yr^{-1}.
\end{equation}
Depending on the observations of pulsating white dwarf star G 117-B
15A, Benvenuto et al. \cite{Benvenuto} have set up the
astroseismological bound as,
\begin{equation}\label{eq:gvar2}
-2.50 \times 10^{-10} \leq \left|\frac{\dot{G}}{G} \right|\leq 4 \times 10^{-10} yr^{-1}.
\end{equation}

\section*{\large{Single component Universe models}}
\begin{figure}[h!]
 \begin{center}$
 \begin{array}{cccc}
\includegraphics[width=50 mm]{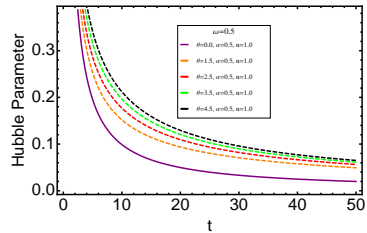}
 \end{array}$
 \end{center}
\caption{Behavior of $H$ against $t$ for a Universe with a barotropic fluid.}
 \label{fig:1}
\end{figure}

\begin{figure}[h!]
 \begin{center}$
 \begin{array}{cccc}
\includegraphics[width=50 mm]{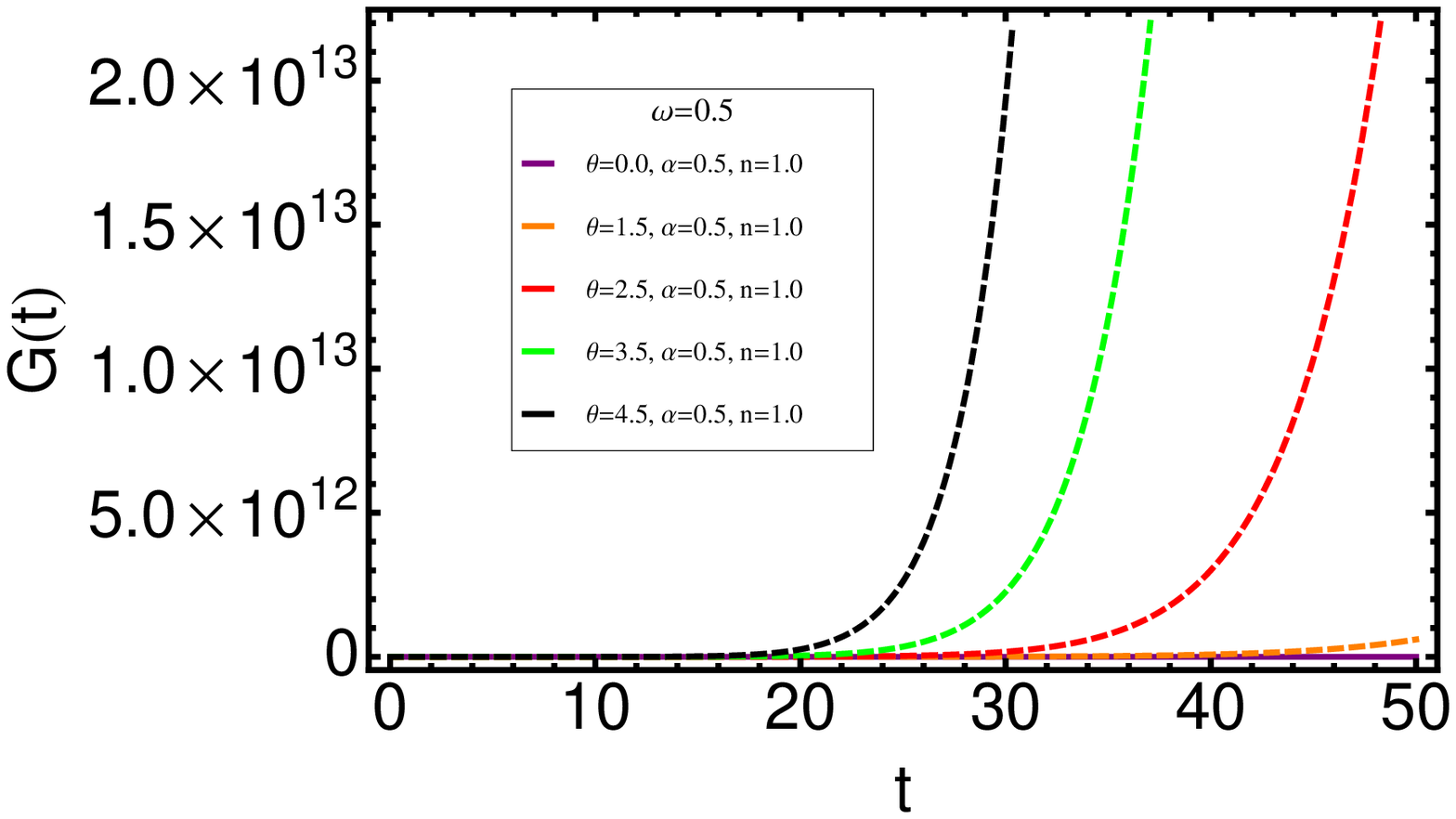}
 \end{array}$
 \end{center}
\caption{Behavior of $G$ against $t$ for a Universe with a barotropic fluid.}
 \label{fig:2}
\end{figure}

\begin{figure}[h!]
 \begin{center}$
 \begin{array}{cccc}
\includegraphics[width=50 mm]{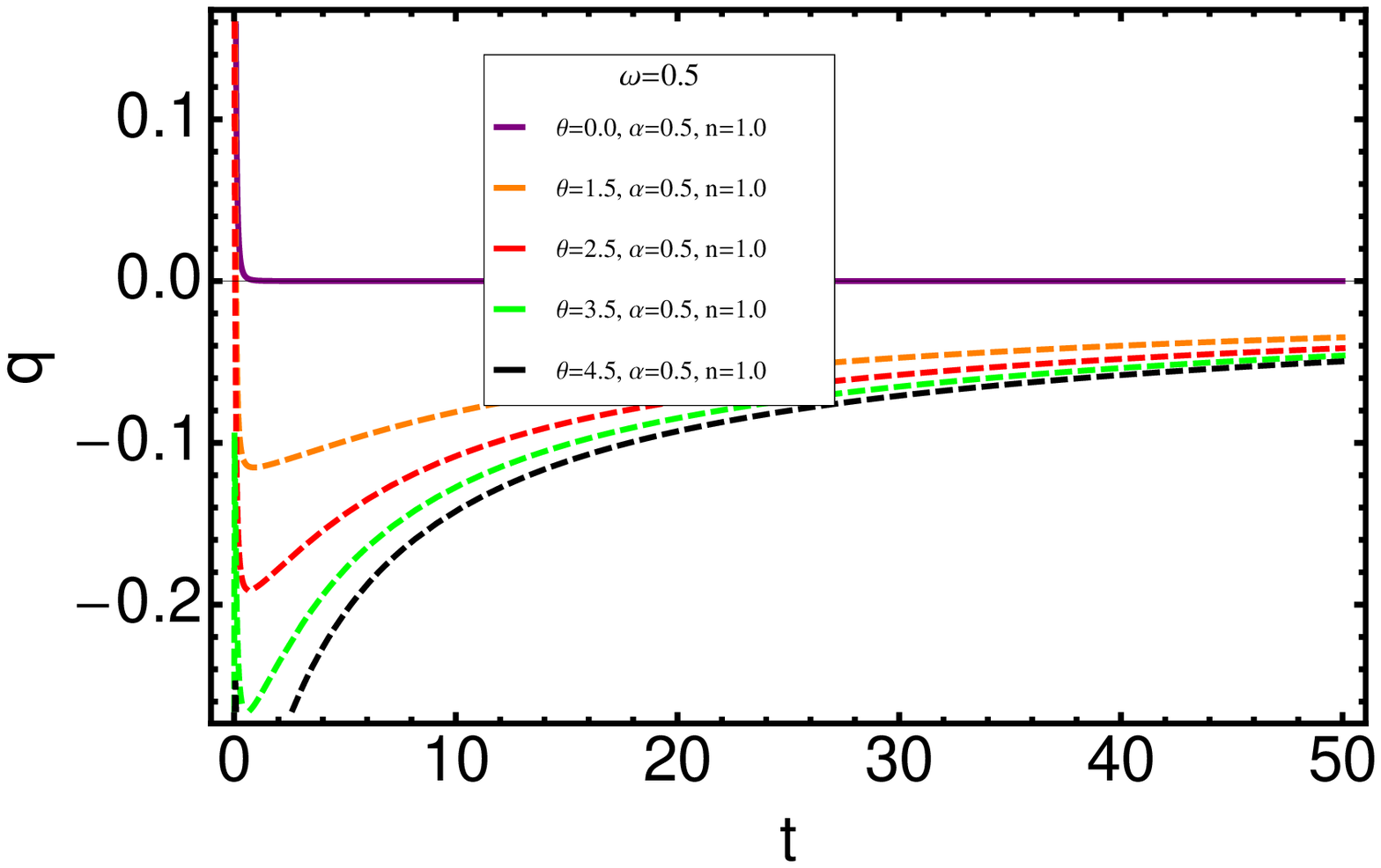}\includegraphics[width=50 mm]{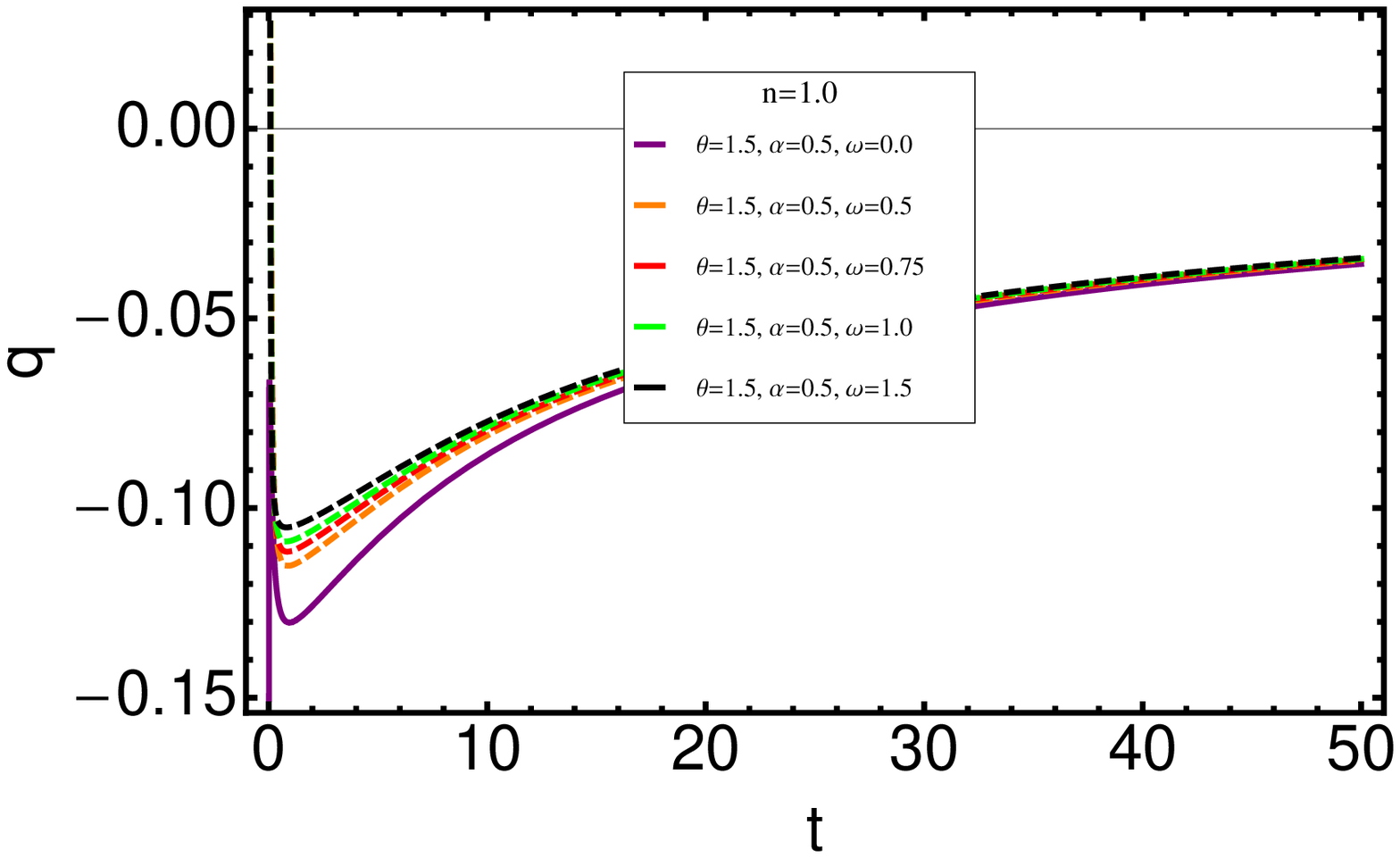}
 \end{array}$
 \end{center}
\caption{Behavior of $q$ against $t$ for a Universe with a barotropic fluid.}
 \label{fig:3}
\end{figure}

With the proposed form of $\Lambda(t)$ we would like to start our investigation of the model and cosmological parameters with other assumption i.e. assuming that our Universe is a one component fluid Universe. Note to complicate our task we will consider two different cases, which are very well studied in literature.\\ One of the EoS for the fluids will correspond to a barotropic fluid of general form given by,
\begin{equation}\label{12}
P_{b}=\omega \rho_{b}^{n},
\end{equation}
with constant EoS parameter $\omega$ and $n>0$. $n=1$ correspond to a usual barotropic fluid case. Behavior of $H$, $G$ and $q$ are given in Fig. \ref{fig:1}-\ref{fig:3}.\\
For example, Fig. 1 shows that increasing parameter $\theta$ increases value of Hubble parameter. As we expected, value of Hubble parameter is decreasing function of time which yields to a constant after long time.\\
In the Fig. 2 we can see that $G$ is increasing function of time, so value of $\theta$ has no important effect at the early Universe. At the late time the parameter $\theta$ increases value of $G$. It is illustrated that the case of $\theta=0$ yields to constant $G$.\\
Plots of the Fig. 3 draw deceleration parameter in terms of time by variation of $\theta$ (left) and $\omega$ (right). They show that increasing $\theta$ decreases $q$ but increasing $\omega$ increases $q$. We confirmed that $q>-1$ and yields to a constant at the late time. Recovering current value of the deceleration parameter need to choose higher value of $\theta$.\\
We also find that variation of $\alpha$ has no important effects on cosmological parameters.

\begin{figure}[h!]
 \begin{center}$
 \begin{array}{cccc}
\includegraphics[width=50 mm]{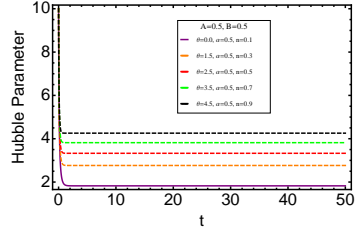}
 \end{array}$
 \end{center}
\caption{Behavior of $H$ against $t$ for a Universe with MCG.}
 \label{fig:4}
\end{figure}

\begin{figure}[h!]
 \begin{center}$
 \begin{array}{cccc}
\includegraphics[width=50 mm]{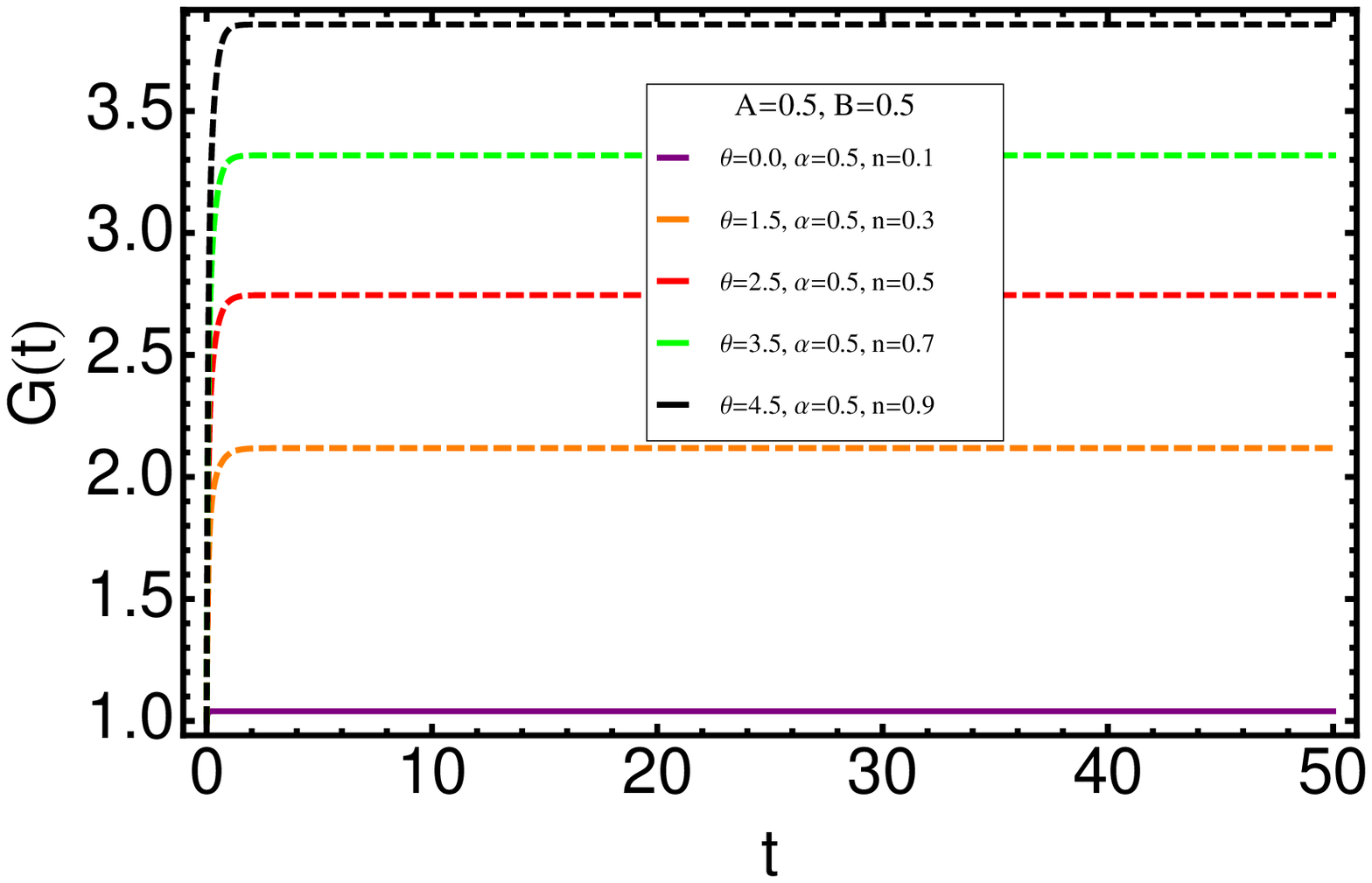}
 \end{array}$
 \end{center}
\caption{Behavior of $G$ against $t$ for a Universe with MCG.}
 \label{fig:5}
\end{figure}

\begin{figure}[h!]
 \begin{center}$
 \begin{array}{cccc}
\includegraphics[width=50 mm]{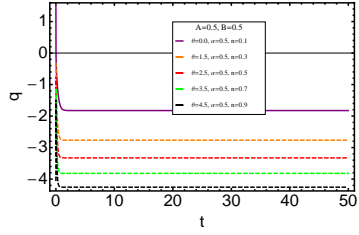}
 \end{array}$
 \end{center}
\caption{Behavior of $q$ against $t$ for a Universe with MCG.}
 \label{fig:6}
\end{figure}

Second one-component fluid model will be associated with a modified Chaplygin Gas (MCG) fluid with the following EoS,
\begin{equation}\label{eq:Chgas}
P_{MCG}=A\rho_{MCG}-\frac{B}{\rho_{\small{MCG}}^{n}}.
\end{equation}
Behavior of Hubble parameter $H$, $G$, and $q$ as a function of time are presented in Figures from \ref{fig:4} to \ref{fig:6}.

\section*{\large{Two-component fluid Universe}}
In this section we will consider two-component fluids as the basis for the energy sources for the Universe. Two-component fluid  will be described by an effective energy density, pressure and EoS parameter $\omega_{\small{eff}}$ given respectively,
\begin{equation}\label{eq:effrho}
\rho=\rho_{1}+\rho_{2},
\end{equation}
\begin{equation}\label{eq:effp}
P=P_{1}+P_{2},
\end{equation}
and,
\begin{equation}\label{eq:effrho}
\omega_{\small{eff}}=\frac{P_{1}+P_{2}}{\rho_{1}+\rho_{2}}.
\end{equation}
Compared to the models considered in the first part of this work. Here will assume that the second component is a GDE, which were assumed to enter to the questions via $\Lambda$.
After some mathematics with field equations for a dynamics of Hubble parameter we can obtain the following equation,
\begin{equation}\label{eq:Hubble eq}
\mathcal{A}\dot{H}+\frac{3}{2}H^{2}+\mathcal{B}H+\mathcal{C}=0,
\end{equation}
where we defined,
\begin{equation}\label{18}
\mathcal{A}=1-\frac{4\pi G(t)}{3}\frac{\theta}{H},
\end{equation}
\begin{equation}\label{19}
\mathcal{B}=-\frac{1}{2}\theta e^{-tH}-4\pi G(t)\theta(1+b),
\end{equation}
and,
\begin{equation}\label{20}
\mathcal{C}=-\frac{1}{2}\frac{\rho_{b}^{2}}{\rho_{b}+\alpha\theta H}+4\pi G(t) \rho_{b}(\omega\rho_{b}^{n-1}-b),
\end{equation}
for barotropic fluid, and
\begin{equation}\label{21}
\mathcal{C}=-\frac{1}{2}\frac{\rho_{CG}^{2}}{\rho_{CG}+\alpha\theta H}+4\pi G(t) \rho_{CG}(A-\frac{B}{\rho_{CG}^{n+1}}-b),
\end{equation}
for MCG. Interaction between fluid components assumed to be,
\begin{equation}\label{eq:interaction}
Q=3Hb(\rho_{i}+\rho_{GD}),
\end{equation}
and dynamics of energy densities can be determined from,
\begin{equation}
\dot{\rho}_{i}=3H(b\theta H -(1-b)\rho_{i} - P_{i}),
\end{equation}
where $i$ stands for barotropic fluid and MCG  for each model respectively.\\
Figs. 7-9 show evolution of cosmological parameter for the case of interacting barotropic fluid and GDE. We find Hubble parameter increased by $\theta$ but $G$ decreased by $\theta$. Fig. 9 shows that the value of deceleration parameter verified $q>-1$ for lower value of $\theta$ and $n$ for example $\theta=0.1$ and $n=0.75$ yields to $\Lambda$CDM model with $q\rightarrow-1$.

\begin{figure}[h!]
 \begin{center}$
 \begin{array}{cccc}
\includegraphics[width=50 mm]{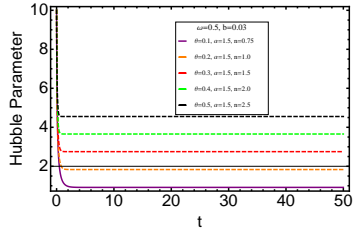}
 \end{array}$
 \end{center}
\caption{Behavior of $H$ against $t$ for a Universe with barotropic fluid and GDE interacting two-component fluid.}
 \label{fig:7}
\end{figure}

\begin{figure}[h!]
 \begin{center}$
 \begin{array}{cccc}
\includegraphics[width=50 mm]{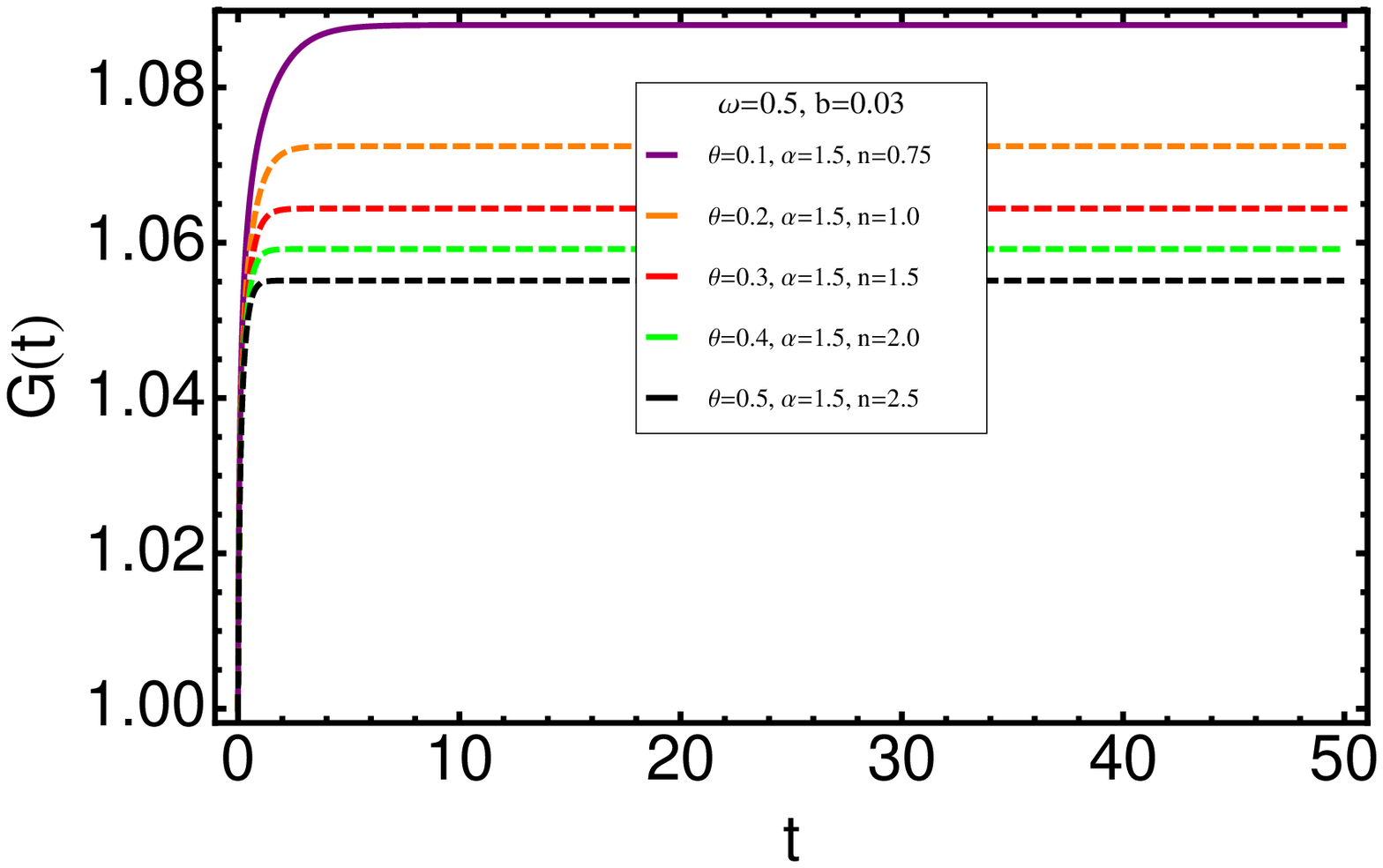}
 \end{array}$
 \end{center}
\caption{Behavior of $G$ against $t$ for a Universe with barotropic fluid and GDE interacting two-component fluid.}
 \label{fig:8}
\end{figure}

\begin{figure}[h!]
 \begin{center}$
 \begin{array}{cccc}
\includegraphics[width=50 mm]{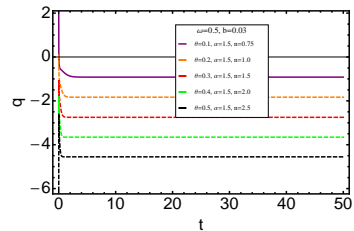}
 \end{array}$
 \end{center}
\caption{Behavior of $q$ against $t$ for a Universe with barotropic fluid and GDE interacting two-component fluid.}
 \label{fig:9}
\end{figure}

The second model is interacting MCG with GDE which yields to the Figs. 10-12. Evolution of Hubble parameter illustrated in the Fig. 10. Evolution of $G$ illustrated in the Fig. 11 and shows that after initial time it yields to a constant. Similar to the previous model observational data suggest lower value of $\theta$ and $n$ for evaluation of $q$ which presented in the Fig. 12.

\begin{figure}[h!]
 \begin{center}$
 \begin{array}{cccc}
\includegraphics[width=50 mm]{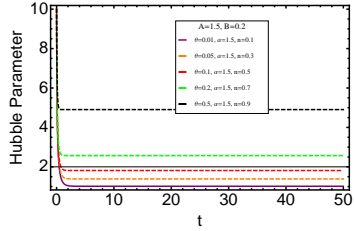}
 \end{array}$
 \end{center}
\caption{Behavior of $H$ against $t$ for a Universe with MCG and GDE interacting two-component fluid with choosing $b=0.02$.}
 \label{fig:10}
\end{figure}

\begin{figure}[h!]
 \begin{center}$
 \begin{array}{cccc}
\includegraphics[width=50 mm]{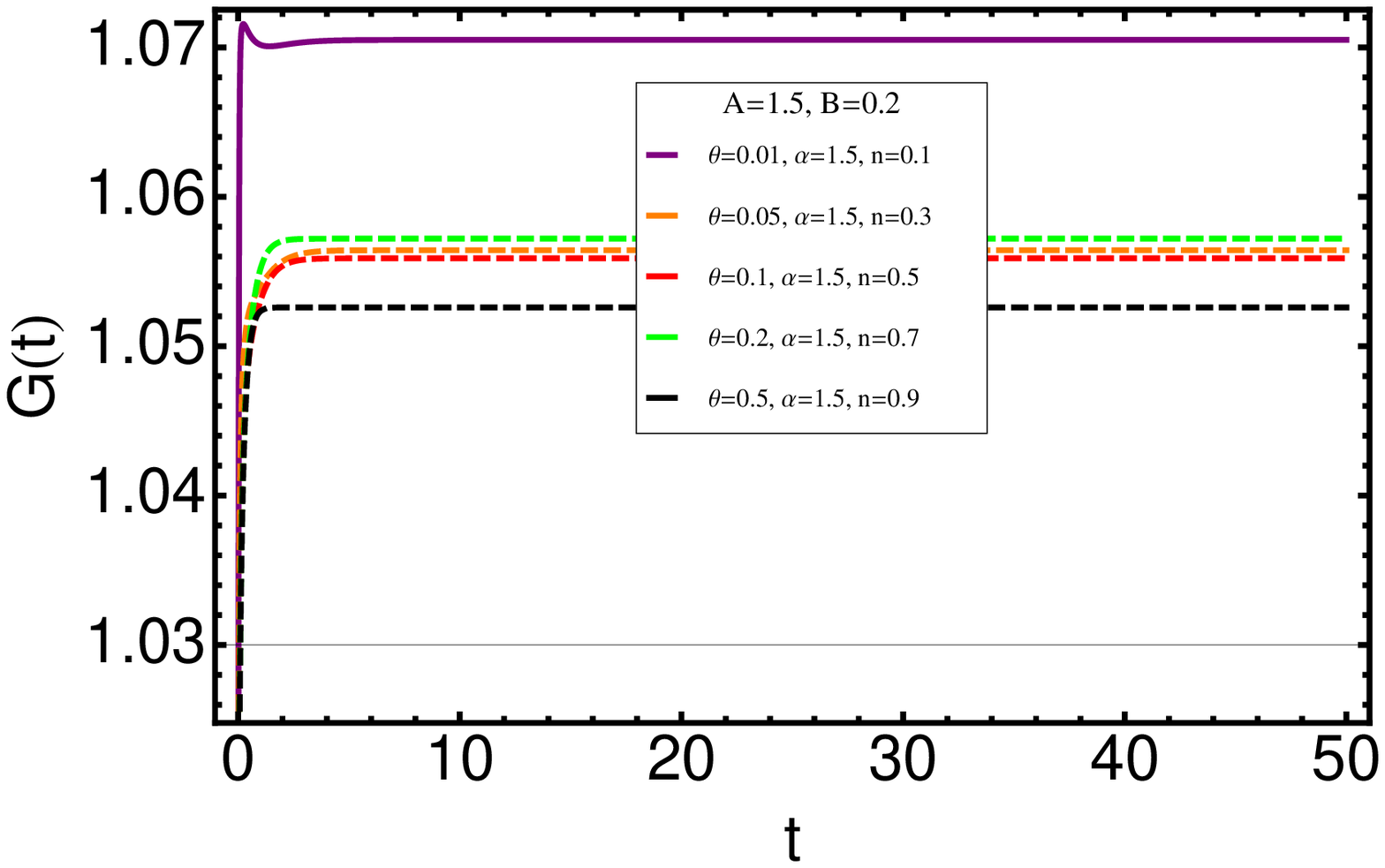}
 \end{array}$
 \end{center}
\caption{Behavior of $G$ against $t$ for a Universe with MCG and GDE interacting two-component fluid with choosing $b=0.02$.}
 \label{fig:11}
\end{figure}

\begin{figure}[h!]
 \begin{center}$
 \begin{array}{cccc}
\includegraphics[width=50 mm]{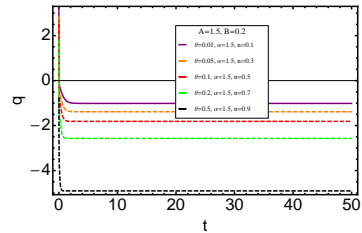}
 \end{array}$
 \end{center}
\caption{Behavior of $q$ against $t$ for a Universe with MCG and GDE interacting two-component fluid with choosing $b=0.02$.}
 \label{fig:12}
\end{figure}

\section*{\large{Discussions}}
In this paper we proposed two toy models for the Universe based on two-component fluid. Also we considered possibility of variable $G$ and $\Lambda$ [32]. In the first model we considered barotropic fluid and GDE as components of Universe which interact with each other. In this case we found Hubble parameter increased by $\theta$ but $G$ decreased by $\theta$. Also we found the value of deceleration parameter for $\theta=0.1$ and $n=0.75$ matches with $\Lambda$CDM observation in which $q\rightarrow-1$.\\
In the second model we considered interacting MCG with GDE and obtain evolution of Hubble parameter, $G$ and deceleration parameter. We found that present value of $G$ should be constant as well as $H$ and $q$.\\
We fixed parameters of model to satisfy generalized second law of thermodynamics [33]. We conclude that the second model (Interacting MCG and GDE) is more coincide with observational data and therefore may be appropriate model to describe Universe. Similar study already performed with another component and varying EoS [34].\\
We also study evolution of total pressure, density and EoS of each model numerically by the plots of the Figs. 13-16.\\
Plots of the Fig. 13 show behavior of total pressure and total energy density of the model of interacting barotropic fluid and GDE. We can see negative constant pressure and positive constant energy density at the late time which increased their value by $\theta$. Also we can see from the Fig. 14 that total EoS yields to -1 at the present epoch.\\
Plots of the Fig. 15 show evolution of the total pressure and total energy density of the model of interacting MCG and GDE. We can see positive pressure at the early universe which suddenly changed to negative constant. This model is also similar to the previous case gives $\omega\rightarrow-1$.

\begin{figure}[h!]
 \begin{center}$
 \begin{array}{cccc}
\includegraphics[width=50 mm]{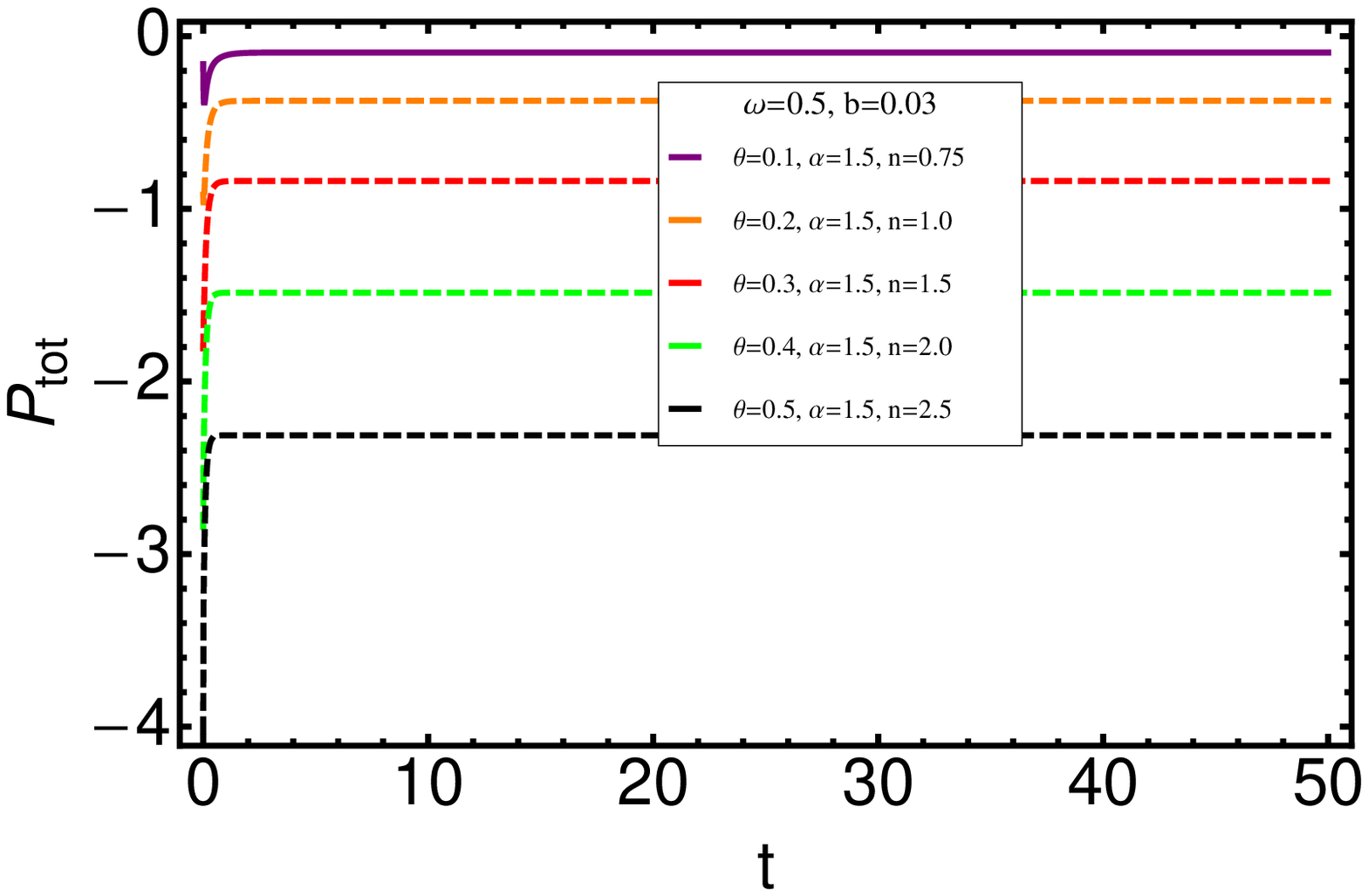}\includegraphics[width=50 mm]{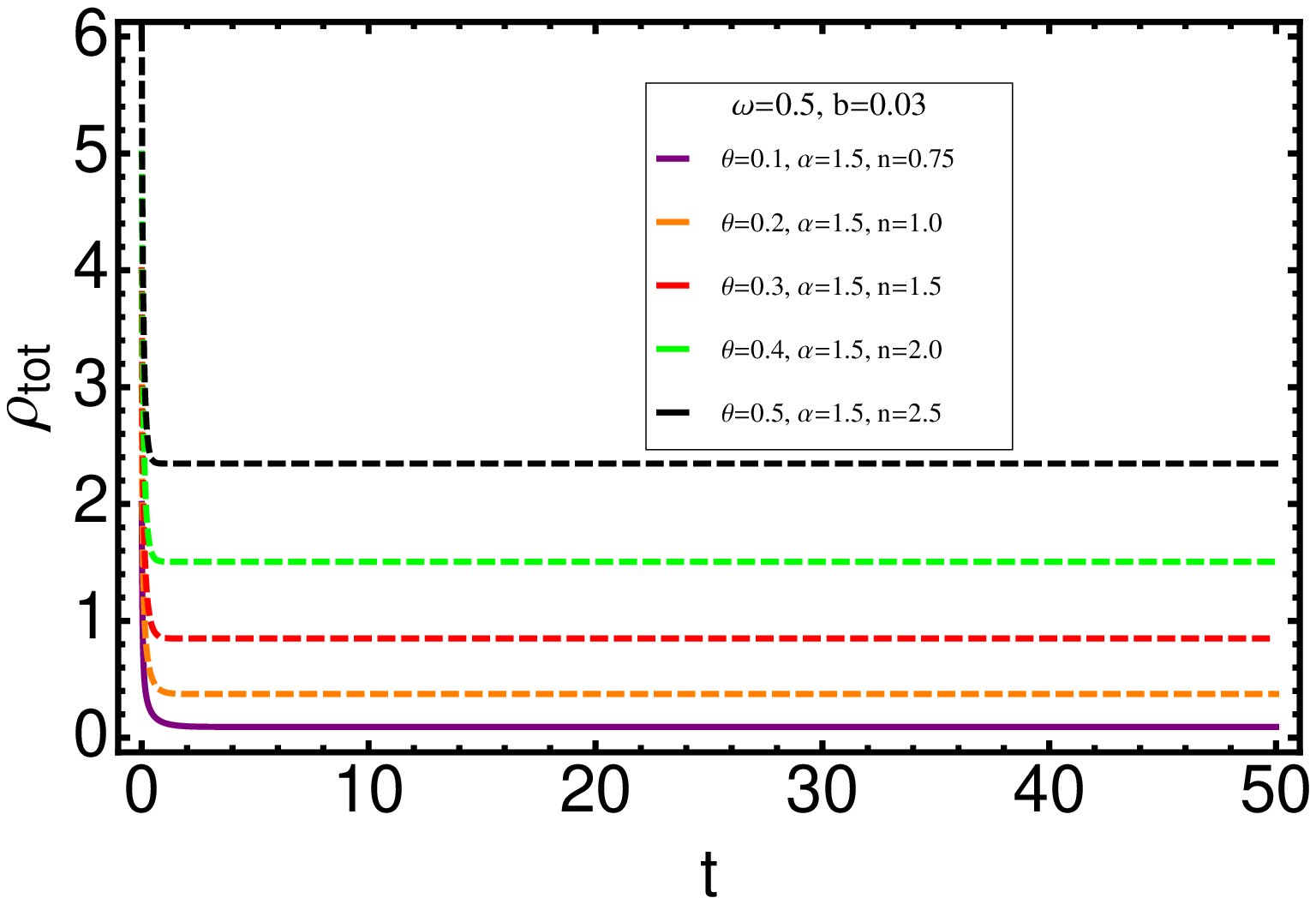}
 \end{array}$
 \end{center}
\caption{Behavior of $P_{tot}$ and $\rho_{tot}$ against $t$ for a Universe with barotropic fluid and GDE interacting two-component fluid.}
 \label{fig:13}
\end{figure}

\begin{figure}[h!]
 \begin{center}$
 \begin{array}{cccc}
\includegraphics[width=50 mm]{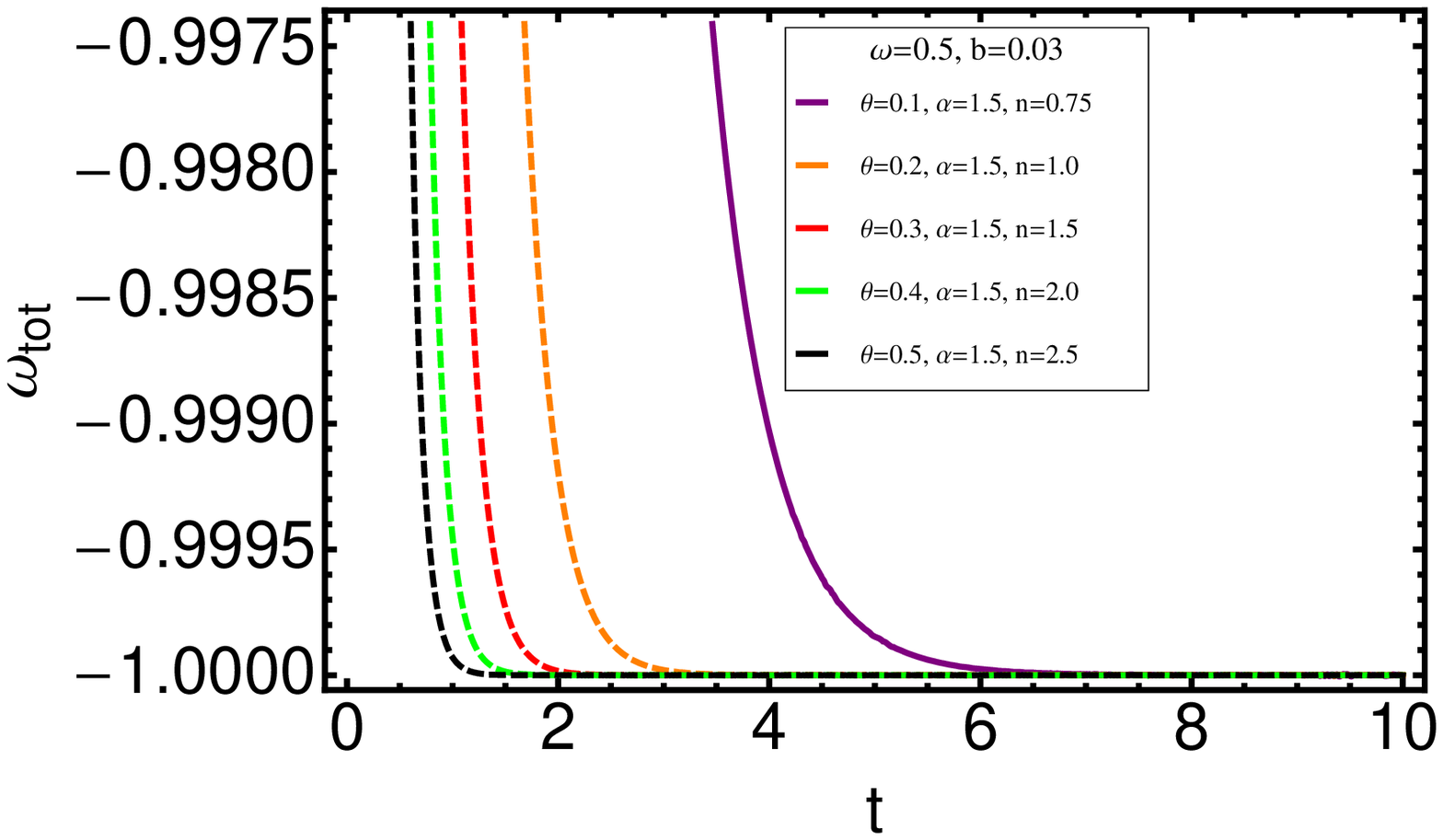}
 \end{array}$
 \end{center}
\caption{Behavior of $\omega_{tot}$ against $t$ for a Universe with barotropic fluid and GDE interacting two-component fluid.}
 \label{fig:14}
\end{figure}

\begin{figure}[h!]
 \begin{center}$
 \begin{array}{cccc}
\includegraphics[width=50 mm]{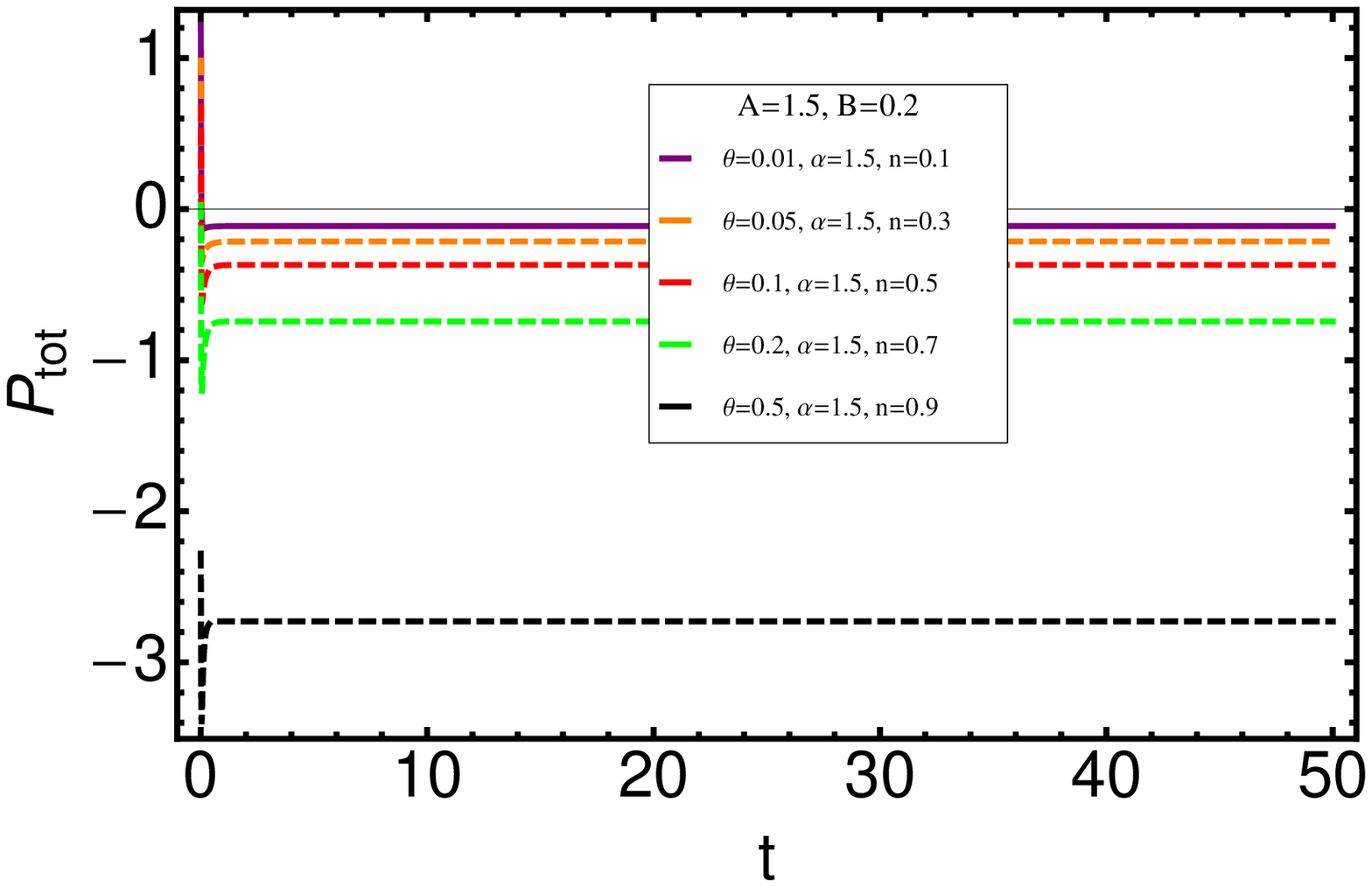}\includegraphics[width=50 mm]{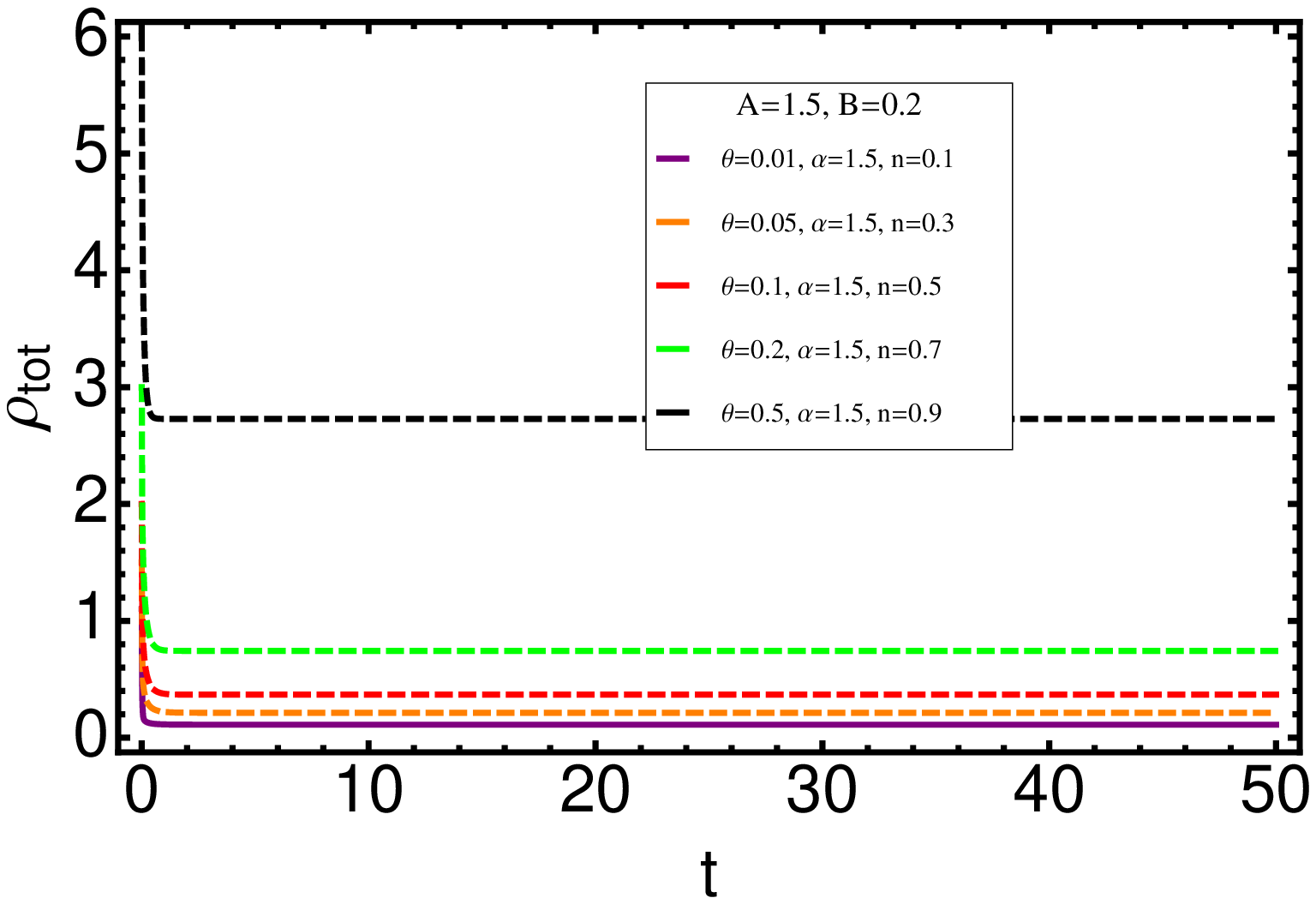}
 \end{array}$
 \end{center}
\caption{Behavior of $P_{tot}$ and $\rho_{tot}$ against $t$ for a Universe with MCG and GDE interacting two-component fluid for $b=0.02$.}
 \label{fig:15}
\end{figure}

\begin{figure}[h!]
 \begin{center}$
 \begin{array}{cccc}
\includegraphics[width=50 mm]{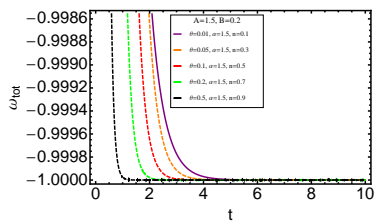}
 \end{array}$
 \end{center}
\caption{Behavior of $\omega_{tot}$ against $t$ for a Universe with MCG and GDE interacting two-component fluid for $b=0.02$.}
 \label{fig:16}
\end{figure}

\end{document}